# A SUSCEPTIBLE-EXPOSED-INFECTED-REMOVED (SEIR) MODEL FOR THE 2009-2010 A/H1N1 EPIDEMIC IN ISTANBUL

Funda Samanlioglu[a], Ayse Humeyra Bilge[b], Onder Ergonul[c]
[a] Department of Industrial Engineering, Kadir Has University, Istanbul, TURKEY.
[b] Faculty of Sciences and Letters, Kadir Has University, Istanbul, TURKEY.
[c] School of Medicine, Koc University, Istanbul, TURKEY.

*Correspondence:* Funda Samanlioglu, Department of Industrial Engineering, Kadir Has University, Central Campus, Kadir Has Caddesi, Cibali 34083, Istanbul, TURKEY. E-mail: fsamanlioglu@khas.edu.tr; Phone: +90 (212) 533 65 32/ 1421; Fax: +90 (212) 533 5753.

**Abstract**

A/H1N1 epidemic data from Istanbul, Turkey during the period June 2009-February 2010 is analyzed with SEIR (Susceptible-Exposed-Infected-Removed) model. The data consist of the daily adult hospitalization numbers and fatalities recorded in various state hospitals serving an adult population of about 1.5-2 million. June 2009-August 2009 period corresponds to the initial stage of the epidemic where the hospitalization rate is nearly %100 and it is excluded from further consideration. The analysis covers the September 2009-February 2010 period, the total number of hospitalizations and fatalities being respectively 869 and 46. It is shown that the maximum correlation between the number of fatalities and hospitalizations occur with a time shift of 9 days and the proportionality constant is $\delta=0.0537$. The SEIR epidemic model is applied to the data by back-shifting the number of fatalities. The determination of the best fitting model is based on the $L_2$ norms of errors between the model and the data and the errors are around %10 and %2.6 for the number of hospitalizations and fatalities, respectively. The parameters in the model are $I_0$, $\eta$, $\varepsilon$ and $\beta$, where $I_0$ is the percentage of people infected initially, $\eta$ and $\varepsilon$ are related to the inverses of the infection and incubation periods and $\beta/\eta$ is the representative of the basic reproduction number. These parameters are determined as $\eta=0.09$ ($1/\eta =11.11 days$), $I_0=10^{-7.4}$, $\varepsilon=0.32$ ($1/\varepsilon =3.125\ days$), $\beta=0.585$, $\beta/\eta=6.5$.

# A SUSCEPTIBLE-EXPOSED-INFECTED-REMOVED (SEIR) MODEL FOR THE 2009-2010 A/H1N1 EPIDEMIC IN ISTANBUL

1. Introduction

The World Health Organization declared the outbreak of A/H1N1 to be a pandemic [1] in June 2009, and then in August 2010 to move into the post-pandemic period with localized outbreaks that are likely continue [2]. Significant amount of data that was collected during this epidemic still constitutes a valuable resource for ongoing research.

Some of the research articles on this specific epidemic are based on clinical research, where they give information about the key parameters of the epidemic such as the duration of the infectious period and the so-called basic reproduction number. In clinical research literature, the basic reproduction number is defined as the mean number of secondary cases induced by an infected individual during the infectious period. The basic reproduction number and other parameters for A/H1N1 have been estimated in USA [13,14,19], Peru [4], India [7], New Zealand [15], and in southern hemisphere countries (Argentina, Brazil, Chile, Australia, New Zealand, and South Africa) [6]. In these papers, the parameters are computed by analyzing epidemic data with statistical techniques. In this study, we used an alternative method, also common in epidemiology, based on the modeling of the spread of the disease with systems of differential equations. The parameters in these differential equations are related to the clinical properties of the disease, but one has to be aware that there is no one-to-one correspondence between the definitions in the mathematical literature and clinical epidemiology literature.

The "Susceptible-Infected-Removed" (SIR) and the "Susceptible-Exposed-Infected-Removed" (SEIR) models constitute the simplest mathematical framework for the description of the spread of diseases where immunity once acquired, cannot be lost. In the SIR model, the individuals in a society are mostly "Susceptible", with usually a few of them being "Infected". The infected individuals transmit the disease to susceptible individuals, and infected individuals either recover or die at a rate proportional to the number of infected people. At any rate, they are "removed" from the pool of susceptible individuals. The terms "Recovered" and "Removed" are both used in the literature. In this paper, the term "Removed" is used since the pool of "Removed" individuals are observed via the number of fatalities. The SIR model has no latent stage; therefore it is inappropriate as a model for diseases with a latent period such as A/H1N1. The incubation period is incorporated to the model as an intermediate stage between "Susceptible" and "Infected" stages leading to the SEIR model. In the SEIR model, individuals are in "Susceptible", "Exposed", "Infected", and "Removed" stages and only one directional passage among these stages are allowed. In the literature, the SEIR epidemic model and its variants have been investigated in several research articles [9-12,16] and used to estimate the epidemiological parameters of measles [17], Ebola [8], and influenza [3,18].

In this research, the A/H1N1 data collected at major hospitals in Istanbul is used after necessary pre-processing. The data were obtained from a detailed study on the 2009-2010 epidemics, and it was in the form of patient reports, including medical reports and the dates of hospitalization and recovery or death [5]. The hospitalization numbers are considered to be representative of the number of infectious people while the fatalities due to the disease are representative of the number of removed individuals, the proportionality constants being called respectively as the "hospitalization rate" and the "death rate". The parameters of the SEIR model can in principle be determined from the time evolution of the fatalities, without the knowledge of the time evolution of the numbers of infected or susceptible individuals. In the SEIR model, it is assumed that the rate of change of the removed individuals is proportional to the number of infected individuals. The data representing the number of infectious and removed individuals were available to test this assumption. It is seen that the standard SEIR model matches with the data only when a delay between the number of infected and removed individuals is introduced. The standard SEIR model is solved numerically, after a back-shift of the number of fatalities and satisfactory match between the theory and the data are determined after introducing a delay of 9 days. The preprocessing of the data and the description of the SEIR model are presented in Sections 2 and 3 along with the application of the SEIR model and the discussion of the results in Sections 4 and 5.

## 2. Preprocessing of the Data

The daily data of adult hospitalized patients and fatalities due to A/H1N1 epidemic in Istanbul is collected during the period of June 2009-February 2010 from major state hospitals in various districts of Istanbul. The population of each district and the hospitals therein are given in Table 1. These figures are needed in order to estimate the total adult population. By considering the fact that the major hospitals accept patients from neighboring districts, it is assumed that the total number of susceptible adult population is approximately 1.5-2 million.

*** Table 1 here*

During the period of June 2009-August 2009, almost all hospitalized patients with A/H1N1 symptoms were people who have recently traveled to Turkey from other countries and they all have been quarantined. Thus, the data for the period of June 2009-August 2009 is excluded from the dynamics of the epidemic model, and the SEIR model is based on the data starting from September 2009. These records include a total of 1039 patients for the June 2009-February 2010 and 869 patients for the September 2009-February 2010 periods with a total of 46 fatalities during the latter one. The graph of the number of hospitalized patients and the cumulative number of fatalities are given in Fig. 1.

*** Fig.1. here*

In the preprocessing of the data, if a patient has died within 15 days of hospitalization, the date of death is assigned as the day of "death", otherwise if he/she has died after a period longer than 15 days, the day of death is assigned as the 15th day after hospitalization, since it is assumed that death after 15 days is due to some clinical complications, and hospital related infections, but not directly due to A/H1N1 infection. The duration of hospitalization for each patient is computed from the data as (day out-day in). The date-in is assigned as the day of "infection". Average duration of hospitalization for all patients is found to be 4.204 days with a standard deviation of 4.028 days, and the average duration of hospitalization for patients who died is found to be 6.609 days with a standard deviation of 5.690 days. From Fig. 1, it can be seen that the number of hospitalizations are quite scattered. This is partly due to the fact that people tend to report to hospitals more after news of fatalities. The number of hospitalizations has thus been smoothed and the original or smoothed data is used as appropriate. In Fig. 1, there is an obvious delay between the peak hospitalization number and the highest rate of change in the fatalities. This leads to the search of an optimal delay between the number of hospitalizations and fatalities based on their correlations. It is found that the highest correlation occurs for 9 day shifts and the slope of the best regression line is $\delta=0.0537$ as shown in Fig. 2.

*** *Fig. 2 here*

In most of the epidemic studies, the most reliable observable data of the systems are the number of fatalities that are proportional to the number of removed individuals. In this study, in addition to the number of fatalities, number of hospitalizations was used as a fraction of the number of infections.

## 3. The SEIR model

In the literature, SEIR models [9-12] refer to the differential equations governing the dynamics of a population where the individuals can be "susceptible (S)", "exposed (E)", "infected (I)" and "removed (R)". The differential equations (1)-(4) of the SEIR model are given as:

$$dS/dt = -\beta I(t) S(t) \qquad (1)$$
$$dE/dt = \beta I(t) S(t) - \varepsilon E(t) \qquad (2)$$
$$dI/dt = \varepsilon E(t) - \eta I(t) \qquad (3)$$
$$dR/dt = \eta I(t) \qquad (4)$$

, where $\beta$, $\varepsilon$ and $\eta$ are constants. Note that the sum $dS/dt+dE/dt+dI/dt+dR/dt=0$, hence $S+E+I+R=N$, where $N$ is the total number of individuals in the population. Here, since $N$ is assumed to be a constant, the effect of normalizing all variables is just a scaling of the parameter $\beta$. The parameters in the model are $I_0$, $\eta$, $\varepsilon$ and $\beta$, where $I_0$ is the percentage of people infected initially, $\eta$ and $\varepsilon$ are related to the inverses of the infection and incubation periods and $\beta/\eta$ is the representative of the basic reproduction number.

This set of equations can be solved by numerical methods, using standard functions of subroutines. The difficulties lie in running the models over a reasonably wide range of parameters and delays.

## 4. Application of the SEIR model

The SEIR epidemic model described in the previous section is applied to the daily data that consists of adult hospitalization numbers and cumulative number of fatalities of Istanbul using both the systems given by Eqns. (1)-(4) after back-shifting the data. To obtain the solution curves that fit the data with the least error, a wide range of parameters is examined. These parameter ranges are $4<\beta/\eta<8$, $0.2<\varepsilon<0.4$, $0.07<\eta<0.12$, and $10^{-8}<I_0<10^{-7}$. The performance of the model, hence the error percentages, is measured by the $L_2$ norm of the difference between the model and the observation. The parameters for the best fitting model are determined as $\eta=0.09$, $1/\eta =11.11$, $I_0=10^{-7.4}$, $\varepsilon=0.32$, $1/\varepsilon =3.125$, $\beta=0.585$, and $\beta/\eta=6.5$, with error percentages %10 and %2.6 for the number of infections and fatalities, respectively. The data and the solution curves obtained by either the ODE45 or the DDE23 packages in Matlab. The solutions for both models are similar and the results obtained from the ODE45 package are shown in Fig. 1. The greater modeling error in the number of infections is due to the fact that the number of hospitalizations are more scattered. It is believed that this spread is due to the variations in the hospitalization ratio (possibly as a result of the panic induced by news on epidemic) rather than the actual number of infections. Also, as the hospitalization rate may vary during different phases of the epidemic, data of number of fatalities is considered to be more reliable than the number of infections and it is concluded that the fit to the number of fatalities is an evidence of the goodness of the model. Hence based on %2.6 modeling error of the number of fatalities, it can be stated that the SEIR system is a satisfactory mathematical model for the observed data. Note that, since the total susceptible population is 1.5-2.0 million, the proportion of people that are infected initially $I_0=10^{-7.4}$ corresponds to less than 1 person that is infected initially. $\varepsilon=0.32$ corresponds to a 3-4 day latent period, whereas $\eta=0.09$ corresponds to an 11-12 day infectious period.

## 5. Conclusions

In this paper, the SEIR epidemic model is used to investigate the health impact of the 2009-2010 A/H1N1 epidemics in Istanbul. Results are based on daily data that consists of number of hospitalized patients and fatalities collected at major hospitals of Istanbul. A 9-day delay between the number of infected and removed individuals is introduced, and the standard SEIR model after a back-shift of the number of fatalities is solved numerically to find the best fitting parameters. For future research, a more detailed analysis of the epidemics in Europe can be done based on daily, and more detailed data for over a period of more than a year. Furthermore, periodicities in the basic reproduction number, effects of vaccination, and variations in the population size can be investigated.

## References

[1] M. Chan, World now at the start of 2009 influenza pandemic, World Health Organization (2009) URL: http://www.who.int/mediacentre/news/statements/2009/h1n1_pandemic_phase6_20090611/en/index.html. (Accessed, 05-11-2012).
[2] M. Chan, H1N1 in post-pandemic period, World Health Organization (2010) URL: http://www.who.int/mediacentre/news/statements/2010/h1n1_vpc_20100810/en/index.html. . (Accessed, 05-11-2012).
[3] G. Chowell, H. Nishiura, L. M. A. Bettencourt, Comparative estimation of the reproduction number for pandemic influenza from daily case notification data, Journal of the Royal Society Interface 4 (2007) 155-166.
[4] G. Chowell, C. Viboud, C. V. Munayco, J. Gomez, L. Simonsen, M. A. Miller, J. Tamerius, V. Fiestas, E. S. Halsey, V. A. Laguna-Torres, Spatial and Temporal Characteristics of the 2009 A/H1N1 Influenza Pandemic in Peru, Plos One 6(6) (2011) Article number e21287.
[5] Ö. Ergönül, S. Alan, Ö Ak, F. Sargın, A. Kantürk, A. Gündüz, Ergin, O. Öncül, İ. İ. Balkan, B. Ceylan, N. Benzonana, S. Yazıcı, F. Şimşek, N. Uzun, A. İnan, G. Eren, M. Ciblak, K. Midilli, S. Badur, S. Gencer, A. İ. Dokucu, Ö. Nazlıcan, S. Özer, N. Özgüneş, T. Yıldırmak, T. Aslan, P. Göktaş, N. Saltoğlu, M. Fincancı, H. Eraksoy, Predictors of fatality for Pandemic H1N1 in Istanbul among hospitalized patients, 15. Congress of Turkish Clinical Microbiology and Infectious Diseases (KLİMİK), 23-27 March 2011, Antalya.
[6] Y. H. Hsieh, Pandemic influenza A (H1N1) during winter influenza season in the southern hemisphere, Influenza and Other Respiratory Viruses, 4(4) (2010) 187-197.
[7] T. Jesan, G. I. Menon, S. Sinha, Epidemiological dynamics of the 2009 influenza A(H1N1) outbreak in India, Current Science 100(7) (2011) 1051-1054.

[8] P. E. Lekone, B. F. Finkenstadt, Statistical Inference in a Stochastic Epidemic SEIR Model with Control Intervention: Ebola as a Case Study, Biometrics 62 (2006) 1170-1177.

[9] G. Li, Z. Jin, Global stability of a SEIR epidemic model with infectious force in latent, infected and immune period, Chaos, Solitons and Fractals 25 (2005) 1177-1184.

[10] M. Y. Li, J. S. Muldowney, Global stability for the SEIR model in epidemiology, Mathematical Biosciences 125(2) (1995) 155-164.

[11] M.Y. Li, J. S. Muldowney, P. Van Den Driessche, Global stability of SEIRS models in epidemiology, Canadian Applied Mathematics Quarterly 7(4) (1999) 409-425.

[12] M. Y. Li, J. R. Graef, L. Wang, J. Karsai, Global dynamics of a SEIR model with varying total population size, Mathematical Biosciences 160 (1999) 191-213.

[13] A. M. Presanis, D. De Angelis, The New York City Swine Flu Investigation Team, A. Hagy, C. Reed, S. Riley, B. S. Cooper, L. Finelli, P. Biedrzycki, M. Lipsitch, The Severity of Pandemic H1N1 Influenza in the United States from April to July 2009: A Bayesian Analysis, PLoS Medicine | www.plosmedicine.org 6(12) (2009) 1-12.

[14] C. Reed, F. J. Angulo, D. L. Swerdlow, M. Lipsitch, M. I. Meltzer, D. Jernigan, L. Finelli, Estimates of the Prevalence of Pandemic (H1N1) 2009 United States April–July 2009, Emerging Infectious Diseases, www.cdc.gov/eid, 15(12) (2009) 2004-2007.

[15] M. G. Roberts, H. Nishiura, Early Estimation of the Reproduction Number in the Presence of Imported Cases: Pandemic Influenza H1N1-2009 in New Zealand, Plos One 6(5) (2011) Article number e17835.

[16] H. Shu, D. Fan, J. Wei, Global Stability of multi-group SEIR epidemic models with distributed delays and nonlinear transmission, Nonlinear Analysis: Real World Applications, 13(4) (2012) 1581-1592.

[17] H. Trottier, P. Philippe, Deterministic Modeling Of Infectious Diseases: Applications To Measles And Other Similar Infections, The Internet Journal of Infectious Diseases 2(1) (2002).

[18] H. J. Wearing, P. Rohani, M. J. Keeling, Appropriate Models for the Management of Infectious Diseases, PLoS Medicine | www.plosmedicine.org 2(7) (2005) e621-e627.

[19] L.F. White, J. Wallinga, L. Finelli, C. Reed, S. Riley, M. Lipsitch, M. Pagano, Estimation of the reproductive number and the serial interval in early phase of the 2009 influenza A⁄H1N1 pandemic in the USA, Influenza and Other Respiratory Viruses 3(6) (2009) 267-276.

**Table 1**
Demographic Data

| District's Name | District's Population | State Hospitals in the District |
|---|---|---|
| Fatih | 455,498 | Haseki, Samatya |
| Sisli | 314,684 | Okmeydani, Sisli Etfal |
| Kadikoy | 550,801 | Haydarpasa, Goztepe |
| Kartal | 427,156 | Kartal |
| Total | 1,748,139 | |

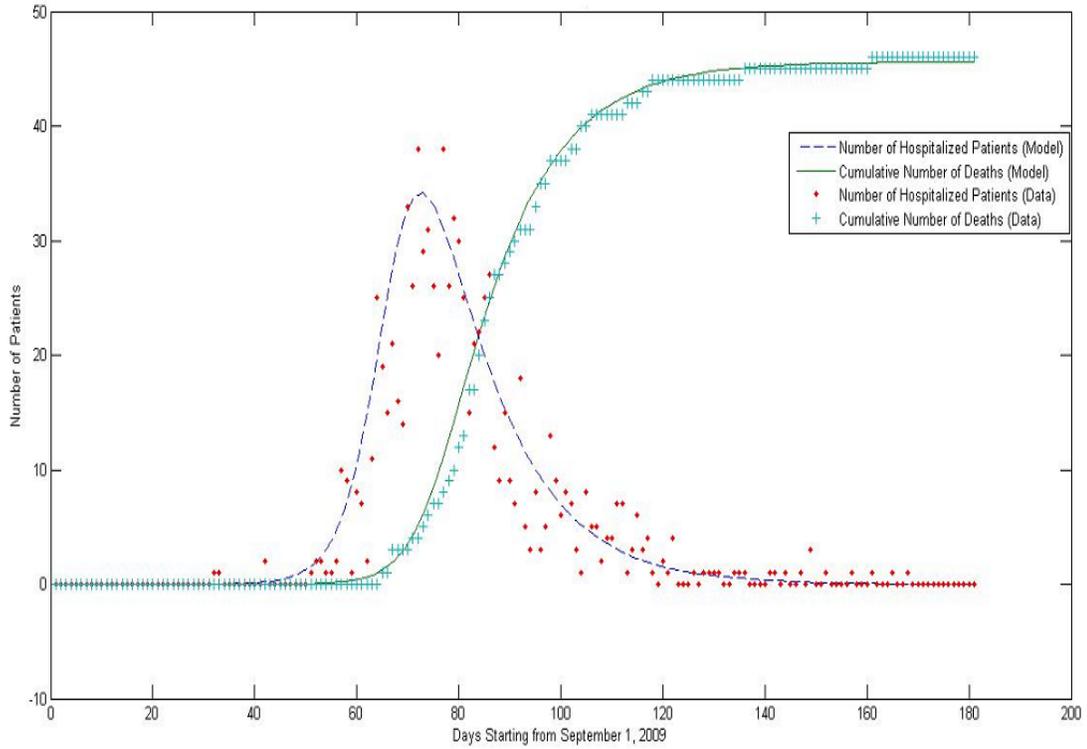

**Fig. 1.** Curves of the SEIR model and daily data

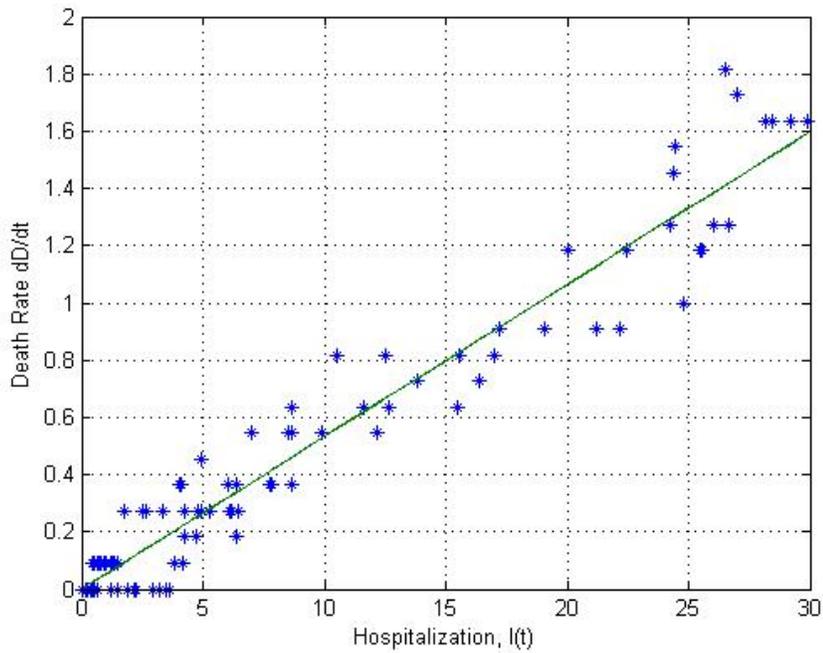

**Fig. 2.** 9-day delay between the number of hospitalizations and fatalities with the slope of the best regression line $\delta$=0.0537.